\documentstyle[12pt,aaspp4,flushrt]{article}
\makeatletter

\newlength{\bibhang}
\let\@internalcite\cite
\def\cite{\@ifstar{\citeyear}{\citefull}}
\def\cite{\let\@citeleft(\let\@citeright)%
    \@ifstar{\citeyear}{\citefull}}
\def\citenp{\let\@citeleft\relax\let\@citeright\relax
    \@ifstar{\citeyear}{\citefull}}
\def\citefull{\def\astroncite##1##2{##1~##2}\@internalcite}
\def\citeyear{\def\astroncite##1##2{##2}\@internalcite}

\def\@citex[#1]#2{\if@filesw\immediate\write\@auxout{\string\citation{#2}}\fi
  \def\@citea{}\@cite{\@for\@citeb:=#2\do
    {\@citea\def\@citea{; }\@ifundefined
       {b@\@citeb}{{\bf ?}\@warning
       {Citation `\@citeb' on page \thepage \space undefined}}%
{\csname b@\@citeb\endcsname}}}{#1}}

\def\@cite#1#2{\@citeleft#1\if@tempswa , #2\fi\@citeright}
\def\@biblabel#1{}

\makeatother
\newcommand{\PSbox}[3]{\mbox{\rule{0in}{#3}\includegraphics{#1}\hspace{#2}}}
\newcommand{\FigNum}[1]{\unitlength 1pt \begin{picture}(55,10)(-400,35) 
                        \put(0,0){Figure #1}
                        \end{picture}}
\newcommand{\msun}{$M_\odot$} 
 
\newcommand{\persec}{\mbox{$\second^{-1}$}}

\newcommand{\cgslum}{\erg\persec}

\newcommand{\approxlt}{\mbox{$\lesssim$}}
\newcommand{\approxgt}{\mbox{$\gtrsim$}}
\def\etal{{et~al.}}

\newcommand{\ee}[1]{\mbox{$10^{#1}$}}
\newcommand{\tee}[1]{\mbox{$\times 10^{#1}$}}

\newcommand{\perval}[2]{{#1\mbox{$^{#2}$}}} 

\def\x1608{{4U~1608$-$522}}

\def\cenx4{{Cen~X$-$4}}

\def\saxj1808{{SAX J1808.4$-$3658}}

\newcommand{\second}{\mbox{$\rm\,s$}}

\newcommand{\erg}{\mbox{$\rm\,erg$}}

\newcommand{\kteff}{$kT_{\rm eff}$}

\newcommand\harding{HL92}

\newcommand\mumag{\mbox{$\mu_{\rm 33}$}}
\newcommand\cmpersec{\mbox{${\rm cm}\;~{\rm s}^{-1}$}}
\newcommand\mmagac{\mbox{$\dot{M}_{\rm MAGAC}$}}
\newcommand\mbh{\mbox{$\dot{M}_{\rm Bondi}$}}
\newcommand\mbondi{\mbox{$\dot{M}_{\rm Bondi}$}}
\newcommand\none{\mbox{$n_{-1}$}}

\slugcomment{\it ApJ; Submitted Dec 13, 2000; accepted Jan 30, 2001}

\begin{document}

\title{Magnetically Accreting Isolated Old Neutron Stars} 

\author{Robert E. Rutledge\altaffilmark{1}} 

\altaffiltext{1}{ Space Radiation Laboratory, California Institute of
Technology, MS 220-47, Pasadena, CA 91125; rutledge@srl.caltech.edu;
and Institute for Theoretical Physics, Kohn Hall, University of
California, Santa Barbara, CA 93106}

\begin{abstract}
Previous work on the emission from isolated old neutron stars (IONS)
accreting the inter-stellar medium (ISM) focussed on gravitational
capture -- Bondi accretion.  We propose a new class of sources which
accrete via magnetic interaction with the ISM.  While for the Bondi
mechanism, the accretion rate \mbondi\ decreases with increasing NS
velocity, in magnetic accretors (MAGACs) \mmagac\ {\em increases} with
increasing NS velocity ($\dot{M}_{\rm Bondi}\propto v^{-3}$
vs. $\dot{M}_{\rm MAGAC}\propto v^{1/3}$).  MAGACs will be produced
among high velocity (\approxgt 100 km \perval{s}{-1}) high magnetic
field ($B>$\ee{14} G) radio pulsars -- the ``magnetars'' -- after they
have evolved first through magnetic dipole spin-down, followed by a
``propeller'' phase (when the object sheds angular momentum on a
timescale \approxlt \ee{10} yr).  The properties of MAGACS may be
summarized thus: dipole magnetic fields of $B$\approxgt \ee{14} G;
minimum velocities relative to the ISM of $>$25--100 km \perval{s}{-1},
depending on $B$, well below the median in the observed radio-pulsar
population; spin-periods of $>$days to years; accretion luminosities
of \ee{28}--\ee{31} \cgslum ; and effective temperatures $kT_{\rm
eff}$=0.3 -- 2.5 keV if they accrete onto the magnetic polar cap.  We
find no examples of MAGACs among previously observed source classes
(anomalous X-ray pulsars, soft-gamma-ray repeaters or known IONS).
However, MAGACs may be more prevelant in flux-limited X-ray catalogs
than their gravitationally accreting counterparts.

\end{abstract}


\section{Introduction}
\label{sec:intro}
Accretion onto isolated old neutron stars (IONS) through the
spherical, Bondi-accretion mode \cite{bondi44,bondi52} has been
investigated as a means to produce a population of neutron stars
detectable in the ROSAT X-ray All Sky Survey (RASS;
\citenp{treves91,blaes93,madau94,zane95,popov00}).  The estimated
number of these sources in the RASS was initially high
($\sim$\ee{3}--\ee{4}).  However, as observations of higher mean
velocities in the radio pulsar population \cite[and references
therein]{lorimer97,hansen97,cordes98} were considered, the predicted
number of detectable IONS accreting from the ISM decreased
dramatically ($\sim$\ee{2}--\ee{3}).  This is because the Bondi
accretion rate is a strong function of the NS velocity through the
interstellar medium (ISM)\footnote{Here, $v=\sqrt{V^2 + C_s^2}$, the
geometric sum of the spatial velocity of the NS and the sound speed in
the accreted medium.  The sound speed of the ISM is low ($\sim$10 km
\perval{s}{-1}) compared to spatial velocities relevant to the present
work ($v$\approxgt 100 km \perval{s}{-1}), so we assume $v\gg C_s$
throughout. } --

\begin{equation}
\dot{M}_{\rm Bondi} =4\pi (GM_{\rm NS})^2 \rho v^{-3}
\end{equation}

\noindent where $\rho$ is the local mass density of the ISM. 
Bondi-accreting IONS are expected to have X-ray luminosities
$\sim$\ee{31} \cgslum\ and effective temperatures $kT_{\rm
eff}\approxlt$ 200 eV, both dependent upon the accretion rate.

There will be a different accretion mode, however, when the NS
magnetic field dominates over the gravitational field in attracting
accretion from the ISM.  In this mode, the warm ionized medium (WIM)
attaches to the NS magnetic field at the magnetosphere, accreting
directly onto the NS surface.  These magnetically accreting isolated
old neutron stars (MAGACs, ``magics'') contrast with IONS which
accrete through the Bondi-mode, in that the Bondi accretion rate
decreases with the NS velocity, while in MAGACs, the accretion rate
{\em increases} with increasing velocity:

\begin{equation}
\dot{M}_{\rm MAGAC} = \pi  R_{\rm M}^2  v  \rho
\end{equation}

\noindent where $R_M=(\mu^2/(4\pi \rho v^2))^{1/6}$ is the size-scale
for the magnetosphere of a non-rotating NS, where the magnetic energy
density from the NS's magnetic moment ($\mu$) is equal to the
ram-pressure of the WIM.  This gives $\dot{M}_{\rm MAGAC} \propto
v^{1/3}$.  So, for a given magnetic field strength, Bondi accretion
dominates at low velocities, while magnetic accretion dominates at
high velocities.

The evolutionary scenario of magnetically accreting NSs has been
outlined previously, by Harding \& Leventhal \cite*[\harding\
hereafter]{harding92}.  \harding\ found magnetised NSs first spin down
via dipole radiation, followed by a longer spin-down period via the
propellor effect.  They restricted their attention to sources with
surface magnetic fields \approxlt \ee{13} G, and found that they are
unable to accrete onto the compact object, as the propellor spin-down
time is longer than the age of the universe.  A similar line of
argument is followed in the review by Treves \etal\ \cite*{treves00}
in the context of Bondi accreting IONS.  This limitation does not
apply to NSs with magnetic fields \approxgt \ee{14} G; a NS with
magnetic field $B=$\ee{15} G and velocity $v$=300 km \perval{s}{-1},
spins down in $\sim$4\tee{9} yr.

On the other hand, this same evolutionary scenario was used to argue
for magnetic field decay in the IONS RX~J0720.4-3125
\cite{wang97,konenkov97}, and for NSs in general
\cite{livio98,colpi98}.  We suggest, instead, that high magnetic field
IONS have not been discovered due to the fact that strong magnetic
field sources accreting from the ISM are spectrally harder (we find
below) than Bondi accretors, and have been selected against by
observers, or confused with background AGN.

Moreover, since \harding's work, observations have revealed NSs with
considerably stronger magnetic fields than considered by \harding, in
the range of \ee{15} G -- the magnetars.  Pulse-timing of
soft-gamma-ray repeaters \cite{chryssa98,chryssa99} confirms their
existence as proposed in earlier theoretical work
\cite{thompson93,duncanthompson92}.  Circumstantial evidence
suggests that anomalous X-ray pulsars have similarly strong $B$-fields
\cite{thompson96,vasisht97,heyl97b}; and precision pulsar-timing
evidence also implies magnetic fields in excess of \ee{14} G
\cite{kaspi99}, confirming less-precise multi-epoch timing solutions
which nonetheless led to a magnetar interpretation \cite{gotthelf99b}.
Static solutions for NSs have been proposed for surface fields up to
$\sim$\ee{18} G \cite{bocquet95,cardall01}, although no objects with
magnetic fields as high as this have yet been observed.  Thus, with
theoretical and observational motivation, we reconsider the scenario
of \harding, but with particular attention paid to NSs with magnetic
fields above \ee{13} G.

In \S\ref{sec:model} we describe the model for this population,
following the scenario of \harding. In \S\ref{sec:discuss}, we
summarize the properties of MAGACs, discuss observational
ramifications and conclude.

\section{Model}
\label{sec:model}

\subsection{Evolutionary Model}

We assume that the neutron star is born with a velocity $v=v_7$\ee{7}
\cmpersec\ relative to the inter-stellar medium, and a magnetic moment
$\mu = \mu_{\rm 33}$ \ee{33} G \perval{cm}{3}, which corresponds
to a dipole field strength at the NS surface (10 km) of \ee{15} G
(appropriate for magnetars; \citenp{thompson93,duncanthompson92}).
The spin period is initially short (P$\ll 10$ sec).  We assume the
magnetic field does not decay.  We assume a compact object mass $M=
M_{\rm NS} 1.4$ \msun, radius $R=R_{\rm NS}$\ee{6} cm.  We assume the
NS travels through a spatially uniform proton density $n_p= 0.1 \none$
\perval{cm}{-3} in the WIM \cite{boulares90,ferriere98a,ferriere98b}.
A more detailed examination of the properties of this population,
which would include a distribution of NS kick velocities, orbits about
a realistic galactic potential, and the galactic gas density and
ionization state -- such as that performed for Bondi accretors and
cooling NSs \cite{popov00,popov01} -- will be the subject of forthcoming work.

First, \harding\ argues that magnetic accretion dominates over Bondi
accretion when the magnetospheric radius $R_{\rm M} = 4.2\times10^{12}
\, \mumag^{1/3} \, v_7 ^{-1/3} \, \none^{-1/6} \; {\rm cm}$ is larger
than the Bondi accretion radius $R_{\rm Bondi} =  \frac{2 G
M_{NS}}{v^2}$. This inequality results in a velocity limitation
for a MAGAC:

\begin{equation}
\label{eq:vlim}
v_{\rm lim} > 94 \; \mumag^{-1/5} \; \none^{1/10} \,
M_{\rm NS}^{3/5} \;  {\rm km\; s^{-1}}
\end{equation}

Electromagnetic dipole radiation pressure inhibits accretion from the
ISM until the light cylinder is external to the magnetospheric radius
($R_{\rm L.C.}>R_{\rm M}$), which occurs at a spin period of

\begin{equation}
P_{\rm L.C.} = 880 \; \mumag^{1/3}\;   v_7^{-1/3} \; \none^{-1/6}\; \;  {\rm s}
\end{equation}

\noindent The timescale to reach this period through dipole radiation
is:

\begin{equation}
\tau_{L.C.} = 2.3\times 10^{7} \; I_{45} \; \mumag^{-4/3} \;
v_7^{-2/3} \; \none^{-1/3} \; \;  {\rm yr}
\end{equation}

\noindent where $I_{45}$ is the moment of inertia in units of \ee{45}
g~\perval{cm}{2}.  This is a short timescale, compared with the life
of the NS.  After reaching this period, the light cylinder is at the
$R_{\rm M}$; however, the NS still cannot accrete (or, accretes
inefficiently) due to the ``propeller effect'' \cite{ill75} in which
matter is spun away by the magnetosphere, which is outside the
Keplerian orbit in co-rotation with the neutron star (the co-rotation
radius).  This matter is ejected from the system, carrying angular
momentum with it, spinning down the NS until the co-rotation radius
reaches $R_{\rm M}$, at which point the NS will have a spin period:

\begin{equation}
P_{\rm P} = 3.9\times 10^{6} \; \mumag^{1/2} \; v_7^{-1/2} \;
\none^{-1/4} \;  M_{\rm NS}^{-1/2} \; \; {\rm s}
\end{equation}

This spin period is substantially longer (by three orders of
magnitude) than any spin period yet observed from a NS.  The propeller
spin-down timescale, when the accretion rate is dominated by
magnetospheric accretion, is \cite{ill75,davies79}:

\begin{equation}
\label{eq:taup}
\tau_{\rm P} = 2.3\times 10^{10} \; I_{45} \; \mumag^{-4/3} \,
v_7^{-5/3} \; \none^{-1/3} \; \; {\rm yr}
\end{equation}

\noindent The $\tau_{\rm P}$ timescale dominates the evolutionary time
between NS birth and emergence of the NS as a magnetic accretor.  It
is comparable to the age of the Galaxy (\approxlt\ee{10} yr)
velocities already observed among the NS population. but only for
high-magnetic field sources (\approxgt \ee{15} G).  This makes
magnetic accretors with the right combination of magnetic field and
velocity observable.

Following spin-down via the propeller, matter is accreted onto onto
the polar cap of the NS.  The maximum accretion rate onto the compact
object is (\harding):

\begin{equation}
\label{eq:mmagac}
\dot{M}_{\rm MAGAC} = \pi  R_{\rm M}^2  v  \rho = 9.6\times 10^{7} \;
\mumag^{2/3} \; v_7^{1/3} \; \none^{2/3} \; \; {\rm g \;   s^{-1}}
\end{equation}

\noindent This $\dot{M}_{\rm MAGAC}$ is an upper-limit to the mass
accretion rate onto the NS surface, and is potentially much lower (see
\S\ref{sec:l}).  Moreover, the Bondi and MAGAC accretion rates are
simply related to the magnetospheric and Bondi radius:

\begin{equation}
\frac{\dot{M}_{\rm MAGAC}}{\dot{M}_{\rm Bondi}} =  \left(\frac{R_{\rm M}}{R_{\rm Bondi}}\right)^2
\end{equation}

If we assume free-fall accretion onto the NS surface, this provides an
accretion luminosity of:

\begin{equation}
\label{eq:lacc}
L_{\rm MAGAC}=1.8\times 10^{28}\; \mumag^{2/3} \; v_7^{1/3} \;
\none^{2/3} \; M_{\rm NS}\; R_{\rm NS}^{-1}\;  {\rm
erg\,  s^{-1}}. 
\end{equation}

\noindent To estimate the effective temperature, we take the size of
the polar cap to be $R_{\rm cap}=R_{\rm NS} (R_{\rm NS}/R_{\rm
M})^{1/2}$.  This would seem unusually small -- $\sim$ 50 m for
$B=$\ee{15} G. Nonetheless, the effective temperature $T_{\rm eff}$ of
this emission is:

\begin{equation}
\label{eq:kteff}
kT_{\rm eff} = 0.39 \; \mumag^{1/4} \; \none^{1/8} \; M_{\rm NS}^{1/4} \; 
R_{\rm NS}^{-1}   \;   \;   {\rm keV}.  
\end{equation}

\noindent Note that $kT_{\rm eff}$ is independent of the NS velocity.
The accretion luminosity will remain below the local Eddington
luminosity if:

\begin{equation}
\frac{\sigma_{\rm es}}{m_H}\, \frac{L_{\rm MAGAC}}{c\,\pi  R_{\rm cap}^2}\, \left(\frac{R_{\rm NS}}{r}\right)^2 
< \frac{G M_{\rm NS}}{r^2}
\end{equation}

\noindent where $m_H$ is the proton mass. The electron scattering
cross-section in a strong magnetic field $\sigma_{\rm es}$ is
suppressed from the Thomson rate $\sigma_T$ by $\sigma_{\rm
es}/\sigma_T=4\times 10^{-6} (E/ 1 {\rm \; keV})^2 (B_{\rm QED}/B)^2$,
with $B_{\rm QED}$=4.4\tee{14} G, and photon energy $E$
\cite{herold79}.  This result gives the constraint $\mumag^{-1}
\none^{1/2} < 3\times10^9$.  This will not be violated for NSs with
$B>$\ee{13} G unless the ISM is un-physically over-dense ($\none\
$\approxgt\ee{6}).  Thus, the accretion rate will be locally
sub-Eddington.

Finally, the ratio of $\dot{M}_{\rm MAGAC}$ to $\dot{M}_{\rm Bondi}$ is: 

\begin{equation}
\label{eq:ratio}
\frac{\dot{M}_{\rm MAGAC}}{\dot{M}_{\rm Bondi}} = 1.8\; \mumag^{2/3}
\; v_7^{10/3} \; \none^{-1/3} M_{\rm NS}^{-2}
\end{equation}

\noindent Thus, for the same velocity distribution, MAGACs will be
more luminous than Bondi accretors.  We discuss the observational
ramifications of this in \S\ref{sec:con}. 

\subsection{Comments on the Accretion Luminosity}
\label{sec:l}
We have two comments on the accretion luminosity.  The first regards
the accretion at magnetopause, and the second regards the need for the
ISM to be ionized in order to be accreted.  

Arons \& Lea \cite*{arons76} performed detailed analytic description
of mass transfer across magnetopause in a \ee{12} G NS system, fed by
a $\approx$600 km \perval{s}{-1} wind from its companion, but with a
substantially higher accretion rate at magnetopause (6\tee{19} g
\perval{s}{-1}) than we consider here.  Even at this lower magnetic
field strength and higher mass accretion rate, they find a cusp forms
with a stand-off shock, causing an interchange instability which
mediates accretion onto the compact object.  While inflow at the cusp
increases with increasing density, the interchange instability
operates more efficiently and prevents very high densities from
arising at magnetopause. Three-dimensional MHD simulations find
similar results \cite{toropin99}.

If the interchange instability of Arons and Lea \cite*{arons76}
permits accretion across the entire NS surface (rather than just at
the magnetic poles), then the luminosity would remain the same, while
dramatically decreasing the effective temperature, to:

\begin{equation}
\label{eq:diffusion}
kT_{\rm eff} = 0.0076 \; \mumag^{1/3}  \; v_7^{1/12}  \; \none^{1/6}
\; M_{\rm
NS}^{1/4}  \; R_{\rm NS}^{-3/4}  \; \; keV. 
\end{equation}

Further analysis of accretion onto a
magnetic dipole with strength \ee{13}--\ee{18} G is necessary to
estimate the magnitude of the effect on the luminosity and spectrum.
For now, we adopt \mmagac\  as an upper-limit.

Secondly, to accrete material via magnetic interaction, the material
must be ionized.  Throughout, we have therefore assumed that MAGACs
accrete from the (ionized) WIM, and not from the neutral ISM.
However, one might wonder if the MAGAC's luminosity is sufficient to
ionize the ISM by the time it reaches the magnetosphere (assuming that
accretion is already taking place).  Ionization of the ISM by an
accreting IONS has been examined previously
\cite{shvartsman71,blaes95}.  These investigations found that a
Bondi-accretion luminosity was capable of ionizing the ISM well beyond
the Bondi-accretion radius.  However, we have applied a similar
approach to MAGACs accreting onto the magnetic polar cap, and find the
fraction of a neutral ISM which would be ionized by the MAGAC at
$R_{\rm M}$ to be $\ll 1$, for a broad range of velocities
(0.1$<v_7<$10) and densities (0.1$<n_{-1}<$100) for magnetic field
strengths considered here.  Therefore, MAGACs must accrete from the
WIM, which has a typical density of $\sim$0.13 \perval{cm}{-3} for a
filling factor of 20\% in the plane and a scale height of 1 kpc
\cite{ferriere98b}.  This justifies our assumption that MAGACs must
accrete from the WIM, and not from the neutral ISM.

\subsection{Limits on the Observability of MAGACs}

Here, we list those effects which will limit the observability of
MAGACs, and which determine the luminosity, spectral hardness,
velocity, magnetic field, and pulse-period parameter space in which
they may be observable.  In Fig.~\ref{fig:bv}, we divide the NS
$B$-field dipole strength, velocity parameter space, on the basis of
these limits:

\begin{enumerate}
\item \mmagac$>$\mbh.  As we find above, this results in a velocity
limit (Eq. ~\ref{eq:vlim}), above which the NS is a MAGAC, and below
which the NS is a Bondi accretor (broken line).

\item {\em The propeller phase must be shorter than the age of the
galaxy}.  We use as a limit (Eq.~\ref{eq:taup}) that
$\tau_p<$1\tee{10} yr (dotted line), and also include the line for
$\tau_p<$1\tee{9} yr.  This age limit also effectively sets a magnetic
field strength limit: NSs with $B<10^{13}$ G will not spin down in a
Hubble time.  Thus, if magnetic fields in NSs decay to $<$\ee{13} G in
less than a Hubble time, they will not be observed as MAGACs.

\item {\em The pulsar must not escape the Galaxy}. We adopt an escape
speed limit of 700 km \perval{s}{-1}
(cf. \citenp{miyamoto75,bohdan90}), although the exact value depends
on where in the galaxy the NS is formed.  This limit is indicated by
the hashed area at $v>$700 km \perval{s}{-1}.  

\item {\em The NS magnetic field should not exceed the stability
limit.}  We assume a surface dipole field of $B<$\ee{18} G
(\citenp{bocquet95,cardall01} ; dashed-dot line).

\end{enumerate}

We see in Fig.~\ref{fig:bv} that acceptable magnetic fields above
\ee{14} G, and acceptable velocities range from $\sim$25 km
\perval{s}{-1} to the escape speed from the galaxy.  We include lines
of constant $L_{\rm MAGAC}$ (for density = 0.1 proton \perval{cm}{-3};
Eq.~\ref{eq:lacc}); the luminosities of MAGACs are between \ee{28} and
\ee{31} \cgslum (again, assuming free-fall accretion from the
magnetosphere onto the NS).  On the right axis of the figure, we mark
$kT_{\rm eff}$, which is independent of velocity, and only weakly
dependent upon the density of the ISM ($\propto n^{1/8}$).  MAGACs
have \kteff=0.3--2.5 keV.  These are harder than observed from any of
the identified IONSs to date \cite{treves00}, perhaps due to
an observational bias.

Finally, we note that for values of $B=$\ee{12} G and $v$=400 km
\perval{s}{-1} which are not unusual in pulsars, such an object will
not accrete via the Bondi-mode from the ISM, as the magnetic field is
in the propeller range.  In addition the source will not spin-down in
$<$\ee{10} yr timescale; so it will wander the Galaxy as a dipole-spun
down, propelling NS.

\section{Discussion and Conclusions}
\label{sec:discuss}
\label{sec:con}

Strong magnetic field ($B>$\ee{14} G) NSs go through three stages of
evolution: (1) dipole emission; (2) propeller slowdown; and finally,
the MAGAC stage (3) magnetic accretion onto the NS surface.  MAGACs
have the following properties: surface magnetic fields between \ee{14}
G up through the NS stability limit ($\sim$\ee{18} G); spin periods
between \ee{6}--\ee{8} s; X-ray luminosities \ee{28}--\ee{31} \cgslum,
for a WIM density of 0.1 protons \perval{cm}{-3}; and effective
temperatures $kT_{\rm eff}$=0.3--2.5 keV.  The low X-ray luminosities of
these sources make it unlikely that the very long spin periods will be
detected as pulsations with present X-ray instrumentation, as
prohibitively long integrations with frequent monitoring would be
required.

At the median velocity for radio pulsars, MAGACs will be
$\sim\times130$ more luminous than Bondi accretors (for $v_7=3$, a
typical median velocity for pulsars; \citenp[and references
therein]{lorimer97,hansen97,cordes98}).  Although the birth-rate of
magnetars has been estimated to be $\sim$10\% of isolated pulsars
\cite{chryssa94,chryssa98}, they can be observed to $\approx$10 times
the distance and $\approx 130\times$ the volume of the Galactic disk
of Bondi accretors at the radio pulsar median velocity.  Magnetic
accretors, therefore, may be more prevalent in flux limited X-ray
catalogs, such as the ROSAT All-Sky-Survey/Bright Source Catalog
(RASS/BSC ; \citenp{voges99}) than the lower-magnetic field Bondi
accretors.  Note, however, that Popov \etal\ \cite{popov00} found that
the detected RASS/BSC INSs are more naturally interpreted as soft,
young, nearby cooling neutron stars, rather than Bondi accretors; in
this interpretation, no NSs accreting from the ISM have yet been
discovered.  The relative number of detectable MAGACs vs. Bondi
accretors will ultimately depend on the particulars of the
distribution of magnetic fields at birth in NSs, the kick velocities
(and their possible dependence on $B$), as well as magnetic field
decay, which we have neglected herein.

A brief comparison between observations and the predicted properties
of this population reveal no strong candidates for MAGACs among known
source classes. 

Among radio pulsars, there are no known objects with implied $B$ and
$v$ which would place them in the MAGAC $B,v$ parameter space.  The
strongest $B$ field radio pulsar is 5.5\tee{13} G \cite{camilo00}, is
just below the low-$B$ limit; its velocity is unmeasured.  At a
velocity of $\sim$ 300 km \perval{s}{-1}, this NS will spin-down on a
timescale of $\tau_{\rm P}\approx$2\ee{11} yr, much longer than the
age of the Galaxy.  Thus, there are no observed radio pulsars in which
the magnetic field is clearly strong enough to evolve into a MAGAC.
This is possibly due to the relatively short timescale that such
objects are dipole emitters ($\tau_{\rm L.C.}$\approxlt\ee{6} yr).

The observed spin periods of magnetars are $P\sim$1-15 sec, which, for
the magnetic fields implied (\ee{15} G), are still within the range
where electromagnetic dipole radiation will inhibit accretion.  In
addition, their X-ray luminosities are a factor \approxgt 1000 larger
than those we expect for MAGACs.  These objects therefore are not
presently MAGACs; however, depending on their spatial velocities, they
may evolve into MAGACs.

The (model dependent) age of the isolated NS RX~J185635-3754 ($\sim$1
My; \citenp{walter96,walter01}) is shorter than $\tau_P$, unless the
NS has close to $B\sim$2\ee{18} G.  With such a high magnetic field,
along with the implied velocity ($\approx$200 km \perval{s}{-1}), the
source would be within the MAGAC parameter space.  Its X-ray
luminosity is $\approx$2\tee{31} \cgslum, which is within a factor of
a few of the expected luminosity for a MAGAC with this $B$ and $v$.
However, the effective temperature ($kT_{\rm eff}= 0.057$ keV is
substantially below that expected from a MAGAC with $B\tee{18} G$ (2.6
keV; Eq.~\ref{eq:kteff}).  The spectral hardness seems to rule out
this object as a MAGAC for the polar-cap accreting scenario.  Since
all previously observed INSs have effective temperatures $<$ 200 eV
\cite{treves00}, it seems similarly unlikely that any known INS is a
MAGAC.  It is possible, however, that an interchange instability at
the magnetosphere permits accretion over a larger fraction of the NS,
which results in a softer spectrum, such as observed here
(ref. \ref{eq:diffusion}).

Discovering these objects in the RASS/BSC requires a similar approach
taken in previous work to identify IONS -- identifying an X-ray source
with no apparent optical counterpart, down to an X-ray-to-optical flux
ratio of $f_x/f_{\rm opt}$\approxgt 1000, with confirmation coming
from an optical detection at an appropriate value. However, MAGACs can
be spectrally harder than their Bondi accreting counterparts ($kT_{\rm
eff}\sim1-3$ keV; cf. \citenp{popov01}), and searches for these
objects must take this into account.  Also, given their strong
magnetic fields, these objects may have a cyclotron {\em emission}
line, which may aid identification through X-ray spectroscopy
\cite{nelson95}.  Moreover, because we have assumed no magnetic field
decay, the absence of MAGACs in the Galaxy, in combination with a
population synthesis with high magnetic field SGRs and AXPs, would
support field decay models.

We conclude that MAGACs may present a population which are equally
prevalent in flux limited X-ray catalogs as Bondi accreting sources.
However, no observed examples of this class are yet known.  

\acknowledgements

I thank Lars Bildsten, Omer Blaes, David Chernoff and Chris Thompson
for useful discussions. I thank Lars Bildsten and Omer Blaes for
comments on the draft of this paper, which improved the manuscript.  I
am grateful to the participants in the ITP Workshop, ``Spin and
Magnetism in Young Neutron Stars'' for stimulating discussions which
brought about this work.  This research was supported in part by the National Science
Foundation under Grant No. PHY99-07949.

\clearpage
\pagestyle{empty}
\begin{figure}[htb]
\caption{ \label{fig:bv} Magnetic field strength ($B$) vs. velocity
for pulsars, showing regions where MAGAC IONS can be observed, and
their accretion luminosities.  The observability of MAGACs are limited
at low velocities (left -- dashed line) by a large magnetosphere
(these then accrete always via Bondi); at high velocities (right --
hashed area) by the likelihood they will escape the Galaxy's
gravitational field; at low $B$ (below -- dotted line) from the fact
that the NSs will not spin down via the propeller mechanism in $<$10
Gyr (we also include a line for $\tau_P=$ 1 Gyr; and at high $B$
(above -- dashed-dot line) by the instability of high $B$ neutron
stars.  Note also that a moderate magnetic field NS (\ee{12} G) with
an average velocity (500 km \perval{s}{-1}) will not accrete
gravitationally due to the propeller mechanism.  We include lines of
the maximum accretion luminosity for the MAGACs after being spun down
via the propeller.  The right axis shows the effective temperature for
accretion onto the magnetic cap, which is independent of velocity and
only weakly dependent on density ($\propto n^{1/8}$). }
\end{figure}

\clearpage
\pagestyle{empty}
\begin{figure}[htb]
\PSbox{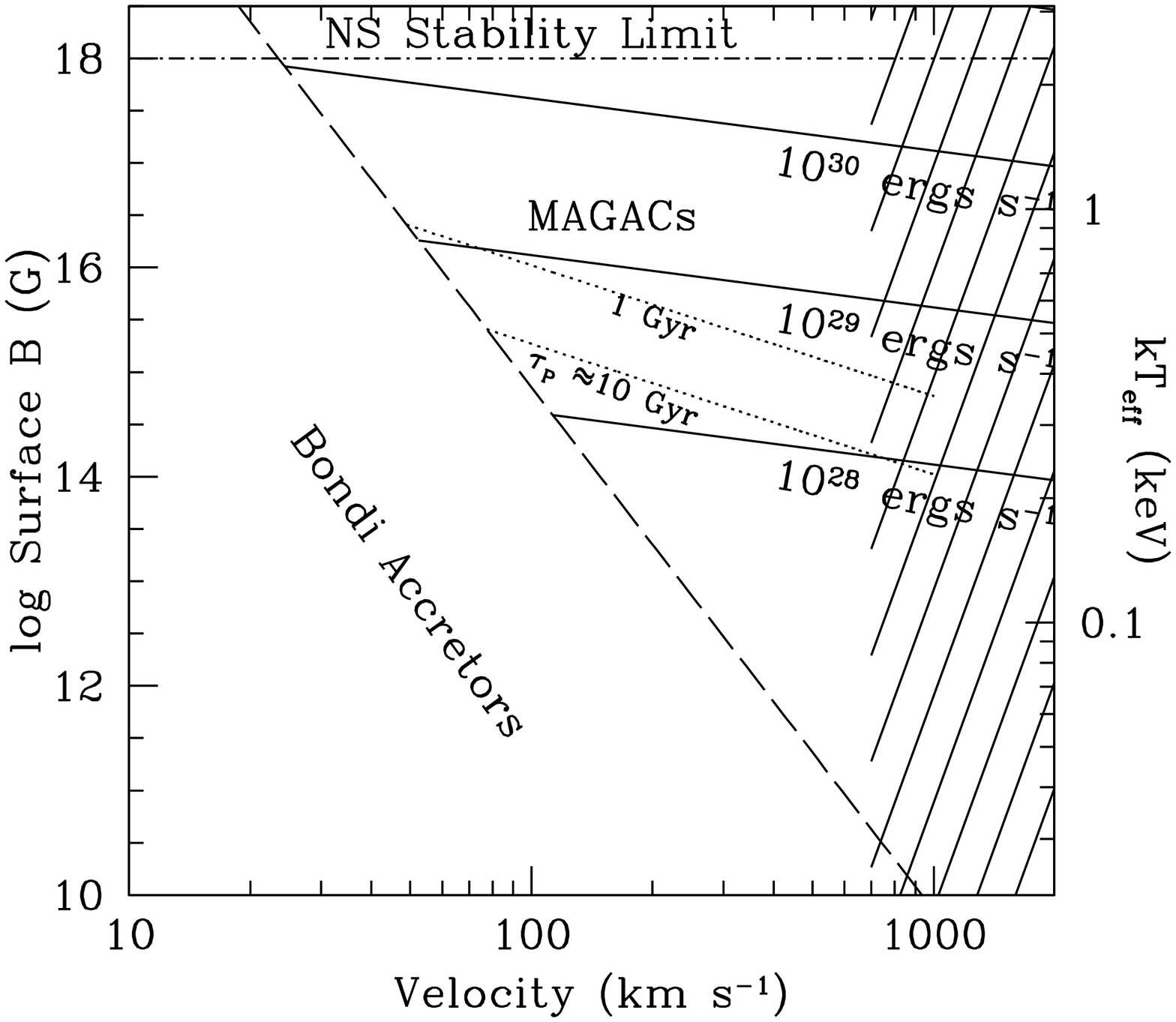 hoffset=-80 voffset=-80}{14.7cm}{21.5cm}
\FigNum{\ref{fig:bv}}
\end{figure}

\end{document}